# Using Google Scholar to Predict Self-Citation: A Case Study in Health Economics

Authors: Richard Norman[1], Francisco M Couto[2]

1. Centre for Health Economics Research and Evaluation (CHERE), University of Technology, Sydney, Australia, 2007. Telephone (+61) 2 95144732; Fax (+61) 2 95144730; Email: Richard.norman@chere.uts.edu.au

2. Department of Informatics, Faculty of Sciences, University of Lisbon

Funding: FC was financially supported through funding of LaSIGE Strategic Project, ref. PEst-OE/EEI/UI0408/2014

Conflict of interests: None

**ABSTRACT**

Metrics designed to quantify the influence of academics are increasingly used and easily estimable, and perhaps the most popular is the h-index. Metrics such as this are however potentially impacted through excessive self-citation. This work explores the issue using a group of researchers working in a well-defined sub-field of economics, namely Health Economics. It then employs self-citation identification software, and identifies the characteristics that best predict self-citation. This provides evidence regarding the scale of self-citation in the field, and the degree to which self-citation impacts on inferences about the relative influence of individual Health Economists. Using data from 545 Health Economists, it suggests self-citation to be associated with the geographical region and longevity of the Health Economist, with early career researchers and researchers from mainland Europe and Australasia self-citing most frequently.

# 1 INTRODUCTION

The *h*-index is perhaps the most popular bibliometric measure of research output. Recent studies considered the merits of this index and its variability across different sources of citation data[1,2]. The *h*-index is sensitive to self-citation in which authors refer to their previous work. While self-citation is often a necessary part of research, it clearly means something different from other citations in terms of evaluating the reach and influence of a researcher's body of publications. Evidence suggests that self-citation differs considerably between disciplines; Thijs and Glänzel[3] report data from 2000-1, in which the share of self-cites ranges from 12% for multidisciplinary researchers to 38% for those in Mathematics.

For individual researchers, self-citation rates will tend to differ over the academic career. Early Career Researchers are likely to have a relatively high self-citation rate for a number of reasons. Firstly, there is likely to be a longer lag time for citations other than self-cites as the broader research community has limited access to work before publication. Secondly, before achieving visibility within the community, it is likely that the number of total cites for a researcher is low. As researchers develop, the possibility of self-citation is likely to increase as their productivity increases. However, for longstanding and successful researchers, while the number of self-cites might be high, the proportion of self-cites may fall as the absolute number of citations increases. The aim of this work is to provide a case study of citation and self-citation, using a sample of researchers working in a specific discipline (Health Economics).

The paper is structured as follows. Firstly, we describe the derivation of the pool of Health Economists which form the analysis set, and outline the method for identifying self-citations in their research. Secondly, we present summary statistics relating to the *h*-index and rates of self-citation in our dataset. Thirdly, we identify the individual researcher characteristics that best predict the rate of self-citations.

**2 MATERIALS AND METHODS**

Google Scholar is a widely used search engine indexing academic literature covering a range of mediums, including peer-reviewed journal articles, technical reports, theses, and books. It is based on the general Google search algorithm, and is gaining traction in the academic community[4]. To have a visible profile on Google Scholar, a researcher registers, and specifies that the profile should be public. At time of analysis, there were 682 Health Economists in this group. The reason we selected Google Scholar over other competing software (such as Web of Science or Scopus) was that it allows identification of researchers who consider themselves part of a specialty (as documented below). The approach to identifying self-citations through Google Scholar is described elsewhere[5]. Briefly, a web tool (http://cids.fc.ul.pt/ ) was developed to interrogate individual Google Scholar profiles, and return a range of user statistics discerning self-citations.

To investigate self-citation, we first had to define it. We selected a relatively sensitive measure, in which papers with any common authors are identified as a self-citation; this approach has been widely employed in the bibliometric literature[6, 7]. A weakness of this choice is that it may inflate self-citation rates in those who tend to publish multi-author papers. However, we will allow for this in a sensitivity analysis by controlling for the average number of authors per cited paper.

The field we selected as a case study (Health Economics) is by definition interdisciplinary, and defining who should be considered as working in the area based on objective criteria (e.g. publication in particular journals, having pre-selected keywords in their research output) is likely to be highly subjective. Therefore, the sample was constructed by analysing each Google Scholar profile with Health Economics as a keyword. Keywords are specified by the user to reflect their personal areas of interest. It is inevitable that some included researchers will have Health Economics as one of a large number of interests, and will have a relatively small subset of Health Economics publications. Similarly, there will be researchers who do Health Economics, but have chosen slightly different keywords. As a robustness check, we considered a variety of areas where Health Economists work but might not be classified as Health Economists in Google Scholar (e.g. Health Econometrics, Health Policy). We did not find a substantial pool of additional Health Economists in these areas, and could not identify a convincing approach for refining the pool of Health Economists beyond the individual's definition of themselves. The final and data extraction was undertaken in January and February 2014, and was limited to those Health Economists with a minimum of 20 citations. These were excluded (n=137) as the rate of self-citation would be sensitive to single self-citations.

The first step in analysis was to consider the distribution of $h$-indices by region, and year of first publication. Since there were small numbers of researchers in many countries, we arbitrarily grouped researchers by their primary affiliation listed on their Scholar profile as documented in Table 1.

The second part of the analysis considered the correlates of rates of self-citation. After visual inspection of the relationship between $h$-index and the proportion of cites that are self-cites, a regression was run to identify the key predictors of the relationship. Independent variables were regional affiliation, $h$-index (linear and quadratic), gender, and a grouped variable reporting year of first publication (with the groups used to construct Figure 2). The last of these was eventually dropped from the analysis as it is highly correlated with measures of $h$-index, and including both causes neither to be statistically significant. The results reported here were from a generalised linear model, with the logit link and the binomial family options, with robust standard errors. Selection of this model reflects that the dependent variable is a proportion, and hence bounded at 0 and 1. To allow for the possibility that those who have more co-authors are more likely to self-cite, a sensitivity analysis was run (Model 2), which added the mean number of authors per cited paper to the main analysis.

**3 RESULTS**

In Google Scholar, we identified 545 researchers who had Health Economics as a keyword and had at least 20 citations. A breakdown of descriptive statistics is presented in Table 1.

**Table 1: Sample Description**

| | | | H-index | | Self-Citation Proportion | |
|---|---|---|---|---|---|---|
| | | N | Mean(SD) | Median(IQR) | Mean(SD) | Median(IQR) |
| | All | 545 | 13.1(11.9) | 9(5-17) | 12.6 (9.5) | 10.4(6.2-16.0) |
| Region | North America | 267 | 13.9(12.8) | 9(5-19) | 10.7 (8.1) | 9.5(5.5-12.8) |
| | UK | 79 | 15.4(14.7) | 11(6-19) | 13.0 (9.4) | 10.8(6.0-17.6) |
| | Other Europe | 119 | 11.5(8.9) | 9(4-16) | 15.7 (9.6) | 14.1(9.3-20.7) |
| | Australia / NZ | 46 | 12.0(9.0) | 9.5(5-17) | 16.0 (13.3) | 12.0(6.3-19.1) |
| | Other | 34 | 8.4(6.8) | 6(5-11) | 11.9 (9.9) | 8.5(5.9-15.6) |
| Gender | Male | 391 | 14.0(12.7) | 10(5-18) | 12.5 (9.5) | 10.3(6.1-15.2) |
| | Female | 154 | 10.7(9.1) | 8(4-14) | 13.0 (9.7) | 10.9(6.3-17.5) |
| Year of First Publication | -1980 | 31 | 33.0(22.3) | 24(16-49) | 9.0 (5.7) | 8.3(5.0-12.3) |
| | 1980-1989 | 50 | 21.8(11.8) | 18(15-27) | 10.5 (5.5) | 10.2(6.0-13.8) |
| | 1990-1994 | 49 | 22.7(12.4) | 19(13-31) | 9.8 (5.2) | 9.1(6.1-12.3) |
| | 1995-1999 | 100 | 15.7(8.3) | 14.5(9-22) | 11.1 (7.1) | 9.8(6.1-15.2) |
| | 2000-2004 | 117 | 10.9(6.5) | 10(7-14) | 12.4 (8.2) | 10.1(6.0-17.5) |
| | 2005- | 198 | 5.4(3.2) | 4(3-7) | 15.3 (12.3) | 11.9(6.9-20.0) |

The proportion of Health Economists working in different countries reflects a similar distribution to that described by Wagstaff and Culyer[8] who, considering all Health-related articles in the EconLit database, show the top ten countries in terms of total number of publications to be (in order) the United States, the United Kingdom, Canada, Australia, Netherlands, Germany, Spain, Sweden, Switzerland, and France.

The pattern of *h*-indices (including self-cites) by region are presented in Figure 1. Median *h*-indices for those with a public Google Scholar profile are highest in the United Kingdom, and lowest in the Other category. The highest *h*-indices are observed almost exclusively in North America and the United Kingdom.

**Figure 1: Box Plot of H-indices, by region**

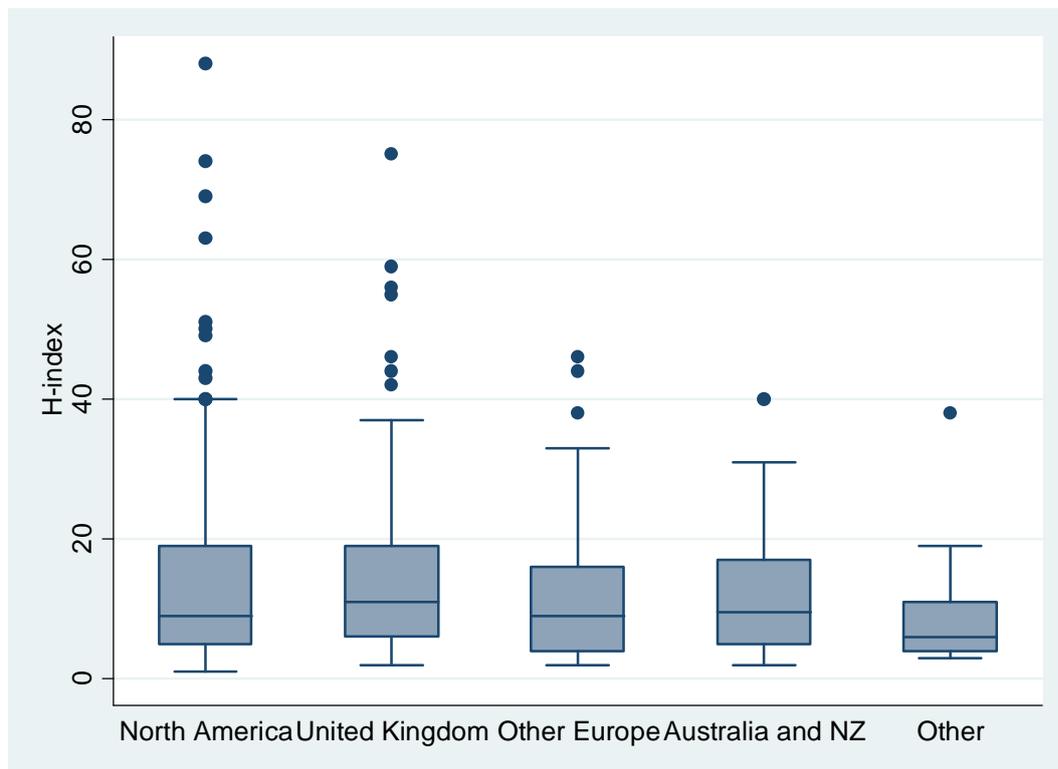

The pattern of *h*-indices by year of first publication is presented in Figure 2. As expected, there is a positive relationship between years of activity and *h*-index; however, it is notable that the gradient is less marked when comparing the relatively older cohorts.

**Figure 2: Box Plot of H-index by Year of First Publication**

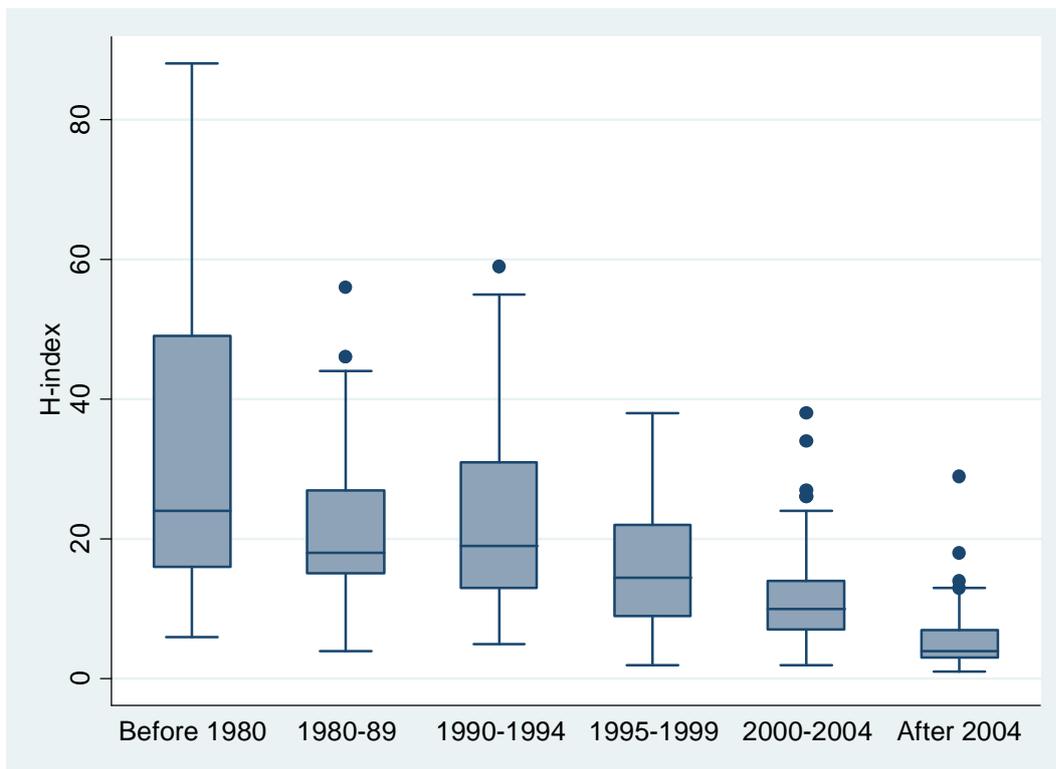

The relationship between *h*-index and self-citation rate is demonstrated in Figure 3. Outliers are either those with low h-indices and high self-citations, or with high *h*-indices, but low self-citation. The correlation co-efficient between the two measures is -0.17, suggesting a weak overall negative relationship.

**Figure 3: Scatter Plot of H-index and Self-Cite Proportion**

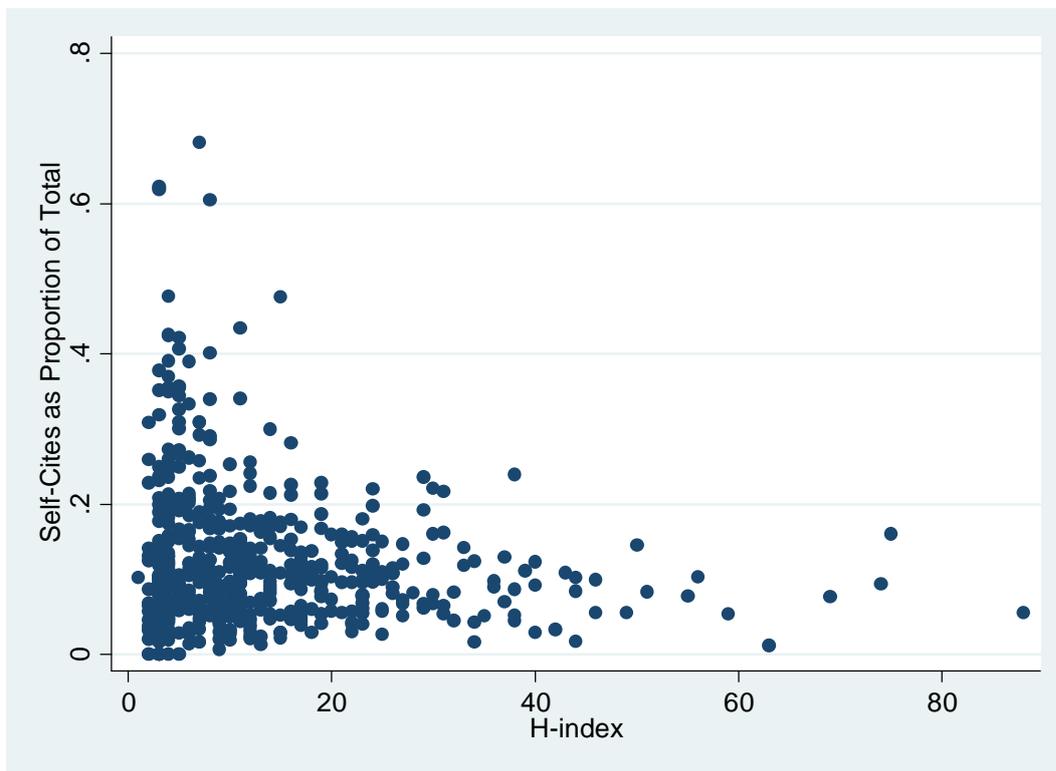

The results from the generalised linear model are presented in Table 2. *H*-index is negatively associated with self-citation rates, but this is partially mitigated by a positive and statistically significant coefficient for the quadratic term. The results for geographic region reflect the descriptive data; relative to the omitted North American group, all other groups self-cite more frequently, although the Other category is not statistically-significantly different. There is no evidence of a gender effect (p=0.702). If the average number of authors per cited paper is included as an additional independent variable, the inferences for the other variables remain largely unchanged despite the statistical significance of the additional coefficient.

**Table 2: Regression analysis**

| Dimension | Independent variable | Model 1 Coefficient (robust SE) | Marginal effect | Model 2 Coefficient (robust SE) | Marginal effect |
|---|---|---|---|---|---|
| H-indices | H-index | -0.023(0.007)*** | -0.003 | -0.023(0.006)*** | -0.003 |
| | (H-index)$^2$/100 | 0.019(0.010)** | 0.002 | 0.018(0.010)* | 0.002 |
| | | | | | |
| Country of affiliation | United Kingdom | 0.240(0.106)** | 0.025 | 0.222(0.109)** | 0.023 |
| | Other Europe | 0.420(0.084)*** | 0.047 | 0.444(0.082)*** | 0.050 |
| | Australia / NZ | 0.461(0.151)*** | 0.053 | 0.423(0.149)*** | 0.047 |
| | Other | 0.055(0.166) | 0.005 | 0.072(0.159) | 0.007 |
| | | | | | |
| Gender | Male | 0.031(0.080) | 0.003 | 0.052(0.079) | 0.006 |
| | | | | | |
| Authors per cited paper | | | | 0.104(0.027)*** | 0.011 |
| | | | | | |
| | Constant | -1.910(0.097)*** | | -2.285(0.121)*** | |

P<0.01 ***; P<0.05 **; P<0.1 *

**4 DISCUSSION AND CONCLUSIONS**

The analysis presented here provides useful descriptive data on the distribution of *h*-indices across different types of Health Economists, and identifies factors which predict self-citation. There is a weak negative correlation between an individual's *h*-index and their rate of self-citation. The region in which someone works is associated with their rate of self-citation, with mainland Europe and Australia and New Zealand displaying relatively high rates, even after controlling for *h*-index. It may be that in countries with fewer Health Economists, the discussion of country-specific topics requires relatively more frequent self-citation.

A key limitation of the analysis is that the sample is based on Health Economists who have a public Google Scholar profile, thus there is the potential for selection bias. It is arguable that those with a public profile are relatively more concerned with citation metrics, and perhaps more likely to self-cite. If this is the case, then the proportion of self-cites can be considered as an upper estimate. Additionally, it is difficult to argue that the selection bias would be of significantly different magnitude across different types of Health Economists, meaning that the relative self-citation rates are not easily explained by this factor.

Are differential patterns of self-citation a major concern? In our data, the correlation coefficient between the standard *h*-index and the *h*-index with self-cites removed is 0.997, suggesting that inferences about the impact of researchers (as defined in this narrow way) are generally not impacted by self-citation. However, as Figure 3 demonstrates, there are sub-groups of researchers with very high self-citation, and judging their research output using an unadjusted *h*-index is problematic. A contribution of this analysis is to provide criteria against which individual self-citation can be benchmarked, something which is of value as citation data becomes an increasingly important part of academic careers.